\documentstyle[12pt]{article}

\newcommand{\be}{\begin{equation}}
\newcommand{\ee}{\end{equation}}
\newcommand{\bea}{\begin{eqnarray}}
\newcommand{\eea}{\end{eqnarray}}

\newcommand{\tilO}{\widetilde{\rm O0}}
\newcommand{\hatO}{\widehat{\rm O0}}

\newcommand{\nbox}{{\,\lower0.9pt\vbox{\hrule \hbox{\vrule height 0.2 cm \hskip
0.2 cm \vrule height 0.2 cm}\hrule}\,}}
\begin{document}
\begin{titlepage}

\begin{flushright}
MIT-CTP-2898\\ TAUP-2596-99\\ RUNHETC-99-30\\ NSF-ITP-99-103\\[1mm]
{\tt hep-th/9909028}
\end{flushright}

\vspace{0.5cm}
\begin{center}

{\Large \bf{Orientifold Points in M Theory}}

\vspace{1.2cm}

Amihay Hanany\footnote{hanany@mit.edu},
Barak Kol\footnote{barak@beauty.tau.ac.il}, and 
Arvind Rajaraman\footnote{arvindra@graviton.rutgers.edu}

\vspace{.6cm}

$^1${\em Center for Theoretical Physics,
\\ Massachusetts Institute of Technology, Cambridge MA 02139.}

\vspace{.3cm}
$^2${\em School of Physics and Astronomy\\
Tel Aviv University, Ramat Aviv 69978, Israel.}

\vspace{.3cm}
$^3${\em Department of Physics and Astronomy,
\\ Rutgers University, Piscataway NJ 08854.}

\vspace{0.5cm}
August 1999

\vspace{.6cm}
\begin{abstract}
We identify the lift to M theory of the four types of orientifold
points, and show that they involve a chiral fermion on an orbifold
fixed circle.
From this lift,
we compute the number of normalizable ground states for the $SO(N)$ and
$Sp(N)$ supersymmetric quantum mechanics with sixteen supercharges.
The results agree with known results obtained
by the mass deformation method.
The mass of the orientifold is identified
with the Casimir energy.
\end{abstract}

\end{center}
\vspace{.5cm}

\end{titlepage}

\section{Introduction}
In this note we shall be considering the supersymmetric matrix quantum mechanics with 16
supercharges. The Lagrangian is
\be
{\cal L}={1 \over 2 g^2} tr \left[  \dot{X}^i \dot{X}^i + 2 \theta^T
\dot{\theta} - {1 \over 2} [X^i,X^j]^2-2\theta^T \gamma_i
[\theta,X^i] \right]
\label{lagrangian}
\ee
where both $X^i, ~1 \le i \le 9$, and the fermionic variables $\theta$,
a nine dimensional spinor, are in the adjoint representation of some Lie algebra $G$.

The case $G=SU(N)$ is the most well studied
 \cite{deWitNicHop,DKPS,KabatPouliot}. This
is the Lagrangian describing $N$ D0 branes in Type IIA
string theory (we shall ignore the decoupled $U(1)$ piece).
The Matrix Theory conjecture \cite{BFSS} asserts that
this Lagrangian, in the large $N$ limit, describes
M theory in the light cone frame.

Since M theory contains a single massless superparticle, there
should be  a normalizable threshold bound state for each value
of $N$ (corresponding to the momentum $N$ graviton).
Hence M
theory and especially Matrix Theory
predict that there should be precisely
one ground state for $G=SU(N)$, i.e. the
Witten index should be 1. This
 prediction was tested and confirmed
by a detailed calculation of the Witten index
\cite{SethiStern,Yi}.

The calculation of the Witten index for an arbitrary Lie
group was performed by Kac and Smilga \cite{KacSmilga}. Their analysis,
to be reviewed in section \ref{MassDeformation}, uses the
mass deformation method, due to Porrati and Rozenberg
\cite{PorratiRozenberg}.
We consider their results for the cases $G=Sp(N), SO(N)$ where they
can be written in terms of a partition function
\be
\sum_{N=1}^\infty d_Nq^N=\prod_r^{\infty}(1+q^r),
\ee
where $r$ runs over even/odd positive integers for $Sp(N/2)$/$SO(N)$ respectively, and
$d_N$ is the supersymmetric index. 
Our goal is to  provide an M theory explanation for the
results of \cite{KacSmilga},
or alternatively, an independent derivation of their result without
making the assumption on which the method of mass deformation hinges.

In section \ref{physical} we provide a realization of this Lagrangian
with $G=Sp(N),$ $SO(N)$ as a system of $N$ D0 branes
of Type IIA
string theory moving in the
background of one of the four types of orientifold points, O0$^+$, O0$^-$,
$\tilO$, and $\hatO$.
We claim that these backgrounds lift to  M theory orbifolds on ${\bf
  R}^9/{\bf Z}_2\times {\bf S}^1$. We argue that the results
of \cite{KacSmilga} can be phenomenologically reproduced
by postulating a chiral fermion living on the
fixed circle (the projection of the fixed circle onto  ${\bf R}^9/{\bf
  Z}_2$ is the fixed point at the origin).
Periodic (R) boundary conditions on the circle correspond to the
$Sp$ groups while anti-periodic (NS) boundary conditions correspond to $SO$ groups.

Independent evidence for this proposal was given by
the results of
Dasgupta and Mukhi \cite{DasguptaMukhi1} , who argued, by an
analysis of gravitational anomalies of M theory on ${\bf T}^9/{\bf Z}_2$ (with (0,16) supersymmetry in
1+1 dimensions),
that the orbifold fixed line carries a chiral fermion in a supersymmetry singlet, in
beautiful agreement with our phenomenology.

It remains to explain how
the different boundary conditions on periodicity R/NS correlate with the
group family $Sp/SO$.
Following a discussion in
\cite{DasguptaMukhi2},
we compare the known masses of the $O0^\pm$
orientifolds with the R/NS Casimir energy in M theory
and find complete agreement.
We find that $O0^+$ and  $\hatO$,
which both give an $Sp(N)$ group, are related through the ground state degeneracy in
the Ramond sector.
 Thus in addition to
rederiving the counting of bound states, we find a physical picture for the various
orientifold points.
 It would be nice to get an M theory picture also for the discrete
 torsions of \cite{WittenAdSBaryons}.

We conclude with remarks on the connection to Matrix  theory in
section \ref{MatrixM}.
Matrix theory compactification  on ${\bf T}^9/{\bf Z}_2$ was also considered in
\cite{rey}, where results consistent with ours
were obtained.

\section{A review of the mass deformation method}
\label{MassDeformation}

In this section we will review the results and arguments of
\cite{KacSmilga}. We can think of the quantum mechanics (\ref{lagrangian}) as a
dimensional reduction from 4 dimensions. Using four dimensional ${\cal N}=1$
 language, the matter content is a vector multiplet with 3 chiral
 multiplets, $\Phi_i,~1 \le i \le 3$ in the adjoint representation, and the
 superpotential is
\be
W=g \epsilon^{ijk} ~f_{ABC} ~\Phi_i^A \Phi_j^B \Phi_k^C
\ee
where $\epsilon$ is the antisymmetric tensor and $f$ are the structure
constants of $G$.
Following \cite{PorratiRozenberg} the superpotential is deformed
by a mass term
\be
\Delta W=M ~\Phi_i^A \Phi_i^A
\ee
The idea is that it is relatively easy to find zero energy solutions
to the deformed Lagrangian. One needs to solve for the D and F
equations, thereby reducing the problem to algebra, and then count the number of
solutions. These vacua break the gauge group completely. For
 large $M$ all degrees of freedom are very massive and
the quantum wavefunction is concentrated around the classical
solution, and so is normalizable. As we take the limit $M \to 0$ these
wavefunctions deform continuously, move towards the center at $\Phi=0$
and become wider. If one assumes (without proof) that all these states
remain normalizable in the limit, then we have a count of the number
of bound states which is easier than a computation of the Witten
index. For $SU(N)$ one gets the
correct result, and our independent derivation for $Sp/SO$ suggests
further that this assumption is valid.

The F and D equations are
\bea
\epsilon_{ijk} ~f^{ABC} ~\Phi_i^A \Phi_j^B \propto M/g ~\Phi_k^C \\
f^{ABC} ~\Phi_i^A \bar{\Phi}_j^B=0.
\eea
To these we add the condition that the vevs break the gauge group
completely, namely that the normalizer in $G$ of the subgroup spanned
by the $\Phi_i$ is trivial. These equations can be interpreted
 \cite{VafaWitten}
to describe embeddings of $sl(2)$ in the
(complexified) group $G$ which have a trivial centralizer.

For the unitary groups there is a unique solution to these equations,
as expected. The solution can be identified with the $N$ dimensional
representation of $SU(2)$ (or $sl(2)$). In general the question can be translated
in mathematical terms to a classification of ``distinguished
markings'' for a Lie algebra. The results are

a) For $SO(N)$ the index is the number of partitions of $N$ into
distinct odd parts.

b) For $Sp(N)$ the index is the number of partitions of $2N$ into
distinct even parts ($Sp(1)=SU(2)$).

The solution corresponding to a given partition can be identified with the
reducible representation of $SU(2)$ with the given partition into
irreducible components.

Anticipating ourselves, we note that
the number of bound states, $d_N$ can be conveniently
coded in a partition function $Z=\sum_{N=1}^{\infty}q^N d_N$
\bea
Z_{SO(N)}= ~\prod_{r=1}^{\infty}(1+q^{2r-1}), \\
Z_{Sp(N)}= ~\prod_{r=1}^{\infty}(1+q^r).
\eea

We realize that phenomenologically, these partition functions
 exactly represent a chiral
fermion on a circle. For $SO$ groups we have half
integer moding (NS), while for $Sp$ it is integral (R).

\section{Physical realization}
\label{physical}
In order to get a physical realization of the results of
\cite{KacSmilga} we should find a physical system for the Lagrangian
(\ref{lagrangian}) with $G=Sp(N), SO(N)$.
 As $Sp(N) \subset SU(2N),
~SO(N) \subset SU(N)$ we start with a system of $N$ D0 branes that
carries an $SU(N)$ group, and use
an orientifold point to project onto the required group.

\subsection{Orientifolds}

Let us review briefly the properties of orientifold planes in
string theory. An orientifold plane (Op plane) is a background for
string theory where we mod out by reversing the coordinates
transverse to the Op plane and reverse the string orientation at
the same time. This background has 16 supersymmetries. To specify
the action on open strings one needs to fix the action on the
Chan-Paton indices, and there are two ways to do that. The two
kinds of O planes are called O$^+$ and O$^-$, and the
corresponding open string gauge groups are $Sp$ and $SO$. An
Op$^\pm$  plane carries RR charge and tension according to $Q=T=
\pm 2^{p-5}$. This is determined by looking at a string diagram on
a test Dp brane.


Actually, one can distinguish a total of 4 kinds of
orientifolds as follows.
From here on we shall specialize to O0's.
(A similar discussion on this point for the case of orientifold
3-planes may be found in \cite{WittenAdSBaryons}.)
The O0 point is surrounded by an ${\bf
 RP}^{8}$.
 We can introduce two ${\bf Z}_2$ valued Wilson lines,
denoted $b$ and $c$, corresponding to nontrivial
gauge configurations of the $B_{NS}$ two-form and
a $C_{RR}$ 5-form (the dual of the 3-form).
The four types of orientifold planes differ in the
values of $(b,c)$.
$b=0$ gives SO groups and $b=1$ gives Sp groups, from the ordinary
coupling of $B_{NS}$ to the worldsheet.

Let us enumerate the different O0 points and their properties. Our
normalization for the RR charge is that a D0 with an image has charge
+1, while a D0 stuck at the orientifold has charge $+{1 \over 2}$.

The O0$^-$ has a trivial value in both ${\bf Z}_2$'s, $(b,c)=(0,0)$,
gives an $SO(2N)$ gauge group
and carries a RR gauge field charge $-{1\over 32}.$

The O0$^+$ has $(b,c)=(1,0)$ and the $\hatO$ has $(1,1)$.
Both carry RR charge $+{1\over 32}$ and give $Sp(N)$ groups.

The $\tilO$ has $(b,c)=(0,1)$, has a D0 stuck on it and thus
carries a RR charge $+{15\over 32}$ and gives $SO(2N+1)$ groups.

In summary, branes which realize an $Sp(N)$ gauge theory
have RR charge $N+{1\over32}$
and branes which realize an $SO(N)$ gauge theory have RR charge
${N\over2}-{1\over32}$.


%

\subsection{A phenomenological fermion}
The Lagrangian (1) with $SO(N)$ and $Sp(N)$ gauge groups
can therefore be realized by $N$ D0 branes at an  orientifold
point. To analyze this system, we need
its M theory interpretation.

When we go from Type IIA to M theory, we have an
extra circle. The orientifold ${\bf R}^9/{\bf Z}_2$ in IIA will
then lift to a space ${\bf R}^9/{\bf Z}_2 \times {\bf S}^1$. We will assume
that the ${\bf Z}_2$ does not act on the ${\bf S}^1$ factor
(this assumption will be justified in the next section, and also by
 the consistency of our analysis).%
Hence there is a singular ${\bf S}^1$ at the origin of the
${\bf R}^9$ coordinates.

Suppose there was a field living on the singular orientifold
circle. This field can then be decomposed into
momentum modes along this circle. A mode with mode
number $N$ along the circle is interpreted in type
IIA as a bound state of $N$ D0 branes sitting at the
orientifold point. We therefore have a bound state for
each allowed mode.

Now suppose that there are a total of $N$ D0 branes in the system.
We can partition this number $N$ into any number of pieces $ N=
n_1+n_2 +\cdots+n_m$. We can then form a state which is a
m-particle state, the first particle being a bound state of $n_1$
D0 branes, the second particle being a bound state of $n_2$ D0
branes,
and so on. 
These states are localized in ${\bf R}^9$ and hence are normalizable. %
Hence it contributes to the Witten index.
The number of ground states is then the number of partitions of
$N$.
\footnote{It is worthwhile to explain why we do not
get an analogous partitioning in the case of
$SU(N)$ gauge theories. The point is that these
bound states in the $SU(N)$ case are interpreted
as 11 dimensional gravitons. A state with
many gravitons has a
relative wave function which is a plane wave and is
therefore not normalizable. In the $SO(N)$ and
$SP(N)$ cases, there is no relative wave function since
all states live at a point in ${\bf R}^9$.}

However, \cite{KacSmilga} found that we actually need the partitions to
be distinct integers. This has a natural interpretation:
the state at the orientifold point should be fermionic.

Furthermore, these states only exist if we have D0 branes
bound to the orientifold point, and not anti-D0 branes,
because a state with an orientifold point and anti-D0 branes is not
supersymmetric. This implies that the fermion can have only
positive mode number i.e. it must be a chiral fermionic
field.

What about the multiplet structure? A standard
D0 brane is a BPS state and breaks 16 of the 32
supersymmetries of Type IIA. These broken supersymmetries
acting on the D0 brane generate a $2^8=256$ dimensional
multiplet of states: the graviton multiplet in
11 dimensions.

Here however, we have a D0 brane along with an orientifold point.
The orientifold point
preserves
the same supersymmetries as
a D0 brane.
However, the
remaining supercharges are not broken, as they can no longer be defined in
the asymptotic space of the orientifold\footnote{ We would like to thank J. Polchinski
for a discussion of this point.}.
There is therefore no multiplet of D0 brane states
in this theory; the D0 brane is a supersymmetry singlet.
(The fact that the D0 brane in type IIA on
${\bf T}^9/{\bf Z}_2$ is a supersymmetry singlet 
was also pointed out in \cite{jatkar}.)

We have therefore been led to postulate a chiral
fermion living on the orientifold singularity.
This fermion is a singlet under the preserved supersymmetries.

The  Hilbert space of
this single chiral fermionic field can be constructed
straightforwardly.

The fermion on the ${\bf S}^1$ can either be in the R or NS
sector i.e. it can be periodic or antiperiodic, respectively.

In the NS sector, the fermion can be expanded into
half-integer modes $b_{n+{1\over 2}}$. These
act on the vacuum denoted by $|0\rangle$.

For total mode number $k
\in {\bf Z}/2$
, the Hilbert space is the set of states given by $b_{n_1+{1\over
2}}b_{n_2+{1\over 2}}\cdots b_{n_m+{1\over 2}}|0\rangle$, with
$k=(n_1+{1\over 2})+(n_2+{1\over 2})+\cdots+(n_m+{1\over 2}) $.
Here the $n_i$ are distinct by the exclusion principle. One should
recall here that the mode number actually counts the number of
physical D0 branes. That is, a state with $n+{1\over2}$ mode
number corresponds to $2n+1$ half physical branes. $k$ can be
either integral or half integral depending on whether $m$ is even
or odd. For the $m$ even case we can rewrite the above equation as
 $N=(2k)=(2n_1+1)+(2n_2+1)+\cdots(2n_m+1)$.
For the $m$ odd case we can rewrite the equation as
$N=(2k+1)=(2n_1+1)+(2n_2+1)+\cdots(2n_m+1)$.

Hence the number of states with $N$ half D0 branes is the number of partitions
of $N$ into distinct odd integers.
Note that the states are always built from an odd number of half D0 branes.

This is identical to the spectrum found by Kac and Smilga for
$SO(N)$ gauge theories.

The R-sector is very similar. The new thing here is that
there is a zero mode $b_0$. Accordingly there are two
vacuum states: $|0\rangle$ and $b_0|0\rangle$.

We can build the Hilbert space by acting by any distinct
combination of $b_n$. The Hilbert space at mode number $k$ is then
the set of states $b_{n_1}b_{n_2}\cdots b_{n_m}|0\rangle$ and
$b_{n_1}b_{n_2}\cdots b_{n_m}b_0|0\rangle$. Here again $n_i$ are
distinct and $k=n_1+n_2 +\cdots n_m$. which we rewrite in physical
D0 brane number as $N=2k=2n_1+2n_2+\cdots 2n_m$.

The number of states in each vacuum sector is then the
number of partitions
of $N$ into distinct even integers.
This is identical to the spectrum found by Kac and Smilga for
$Sp(N)$ gauge theories.

Furthermore, we have two vacuum states. These presumably
correspond to the two types of $Sp(N)$ orientifold planes, the case with no $b_0$ state
corresponds to the familiar $O^+$ point and the case with the $b_0$ state
corresponds to the more exotic $\hatO$ point.

\section{More on M theory}

We can actually find confirmation of these ideas through a
remarkable analysis of Dasgupta and Mukhi \cite{DasguptaMukhi1},
who considered compactifications of M theory on ${\bf T}^9/{\bf
Z}_2$. This theory has (0,16) supersymmetry in 1+1 dimensions.

Let us review their analysis. Eleven dimensional supergravity
contains 128 bosonic degrees of freedom and 128 fermionic
partners. One dimensionally reduces on ${\bf T}^9/{\bf Z}_2$
 and identifies the untwisted sector, which is made of a
gravity multiplet and  128 dimensionally reduced scalars accompanied by
 128 right moving fermionic partners. These scalars and fermions
 are arranged in 16 supermultiplets of (0,16) supersymmetry in 1+1
 dimensions.
This matter content is anomalous. Computation shows that the
gravity multiplet has 3 times the anomaly of the 16 matter
supermultiplets.
In total we need $128\cdot(1+3)=512$ left moving
fermions to cancel the anomaly. This suggests that each orbifold
line has a left moving fermion on it, since there are $2^9$ fixed
points for the orbifold action
(being left moving means that
 it can be excited without destroying the right moving supersymmetry).

This is in exact agreement with what we find.

We can furthermore (following \cite{DasguptaMukhi2}) compute
 the ground state energy of this
compactification. This is nonzero, because of the Casimir
energy of the chiral fermions and chiral bosons.

We need to know for this calculation the Casimir
energy of

a) a periodic boson: $-{1\over 24R}$

b) a periodic fermion: ${1\over 24R}$

c) an antiperiodic fermion: $-{1\over 48R}$

Now the right moving fields are in supermultiplets, and do not
contribute to the Casimir energy. Among the left-movers, in the NS
case, there are 128 periodic bosons and 512 antiperiodic fermions.
The total energy is then $-{16\over R}$. Since there are 512 fixed
points, the energy per fixed point is then $-{1\over 32R}$. A
similar calculation in the R sector yields an energy per fixed
point equal to ${1\over 32R}$.

This is in precise agreement with the masses of
$O0^-$ and $O0^+$ orientifold points. 
Note that there is a bulk contribution to the mass,
$-{1\over96R}$, from the untwisted sector which is common to all
kinds of orientifold points, while the twisted fermion
contribution differs.


One of the corollaries of our construction is
that all the orientifold points must be realized in M theory
as a ${\bf R}^9/{\bf Z}_2 \times {\bf S}^1$ space. For suppose
that the ${\bf Z}_2$ action did not leave the
${\bf S}^1$ invariant. There are two options:
the ${\bf Z}_2$ can act as a shift, or it
can act as a reflection. In the first case, the
resulting space is smooth, with no fixed points, and
there is no natural way to introduce a new field.
In the second case, there
are two singular points, not
a singular line, and we cannot introduce a one-
dimensional field. This only leaves the case that
the ${\bf Z}_2$ does not act on the ${\bf S}^1$.

\section{Matrix theory}
\label{MatrixM}

The preceding discussion
leads to a Matrix theory realization of M theory on the space
${\bf R}^9/{\bf Z}_2$. Explicitly, we find that $SO(N)/Sp(N)$
gauge theories with 16 supercharges describe DLCQ M theory on
${\bf R}^9/{\bf Z}_2$ with a fermion in the $NS/R$ sector
respectively. In this interpretation, the ${\bf S}^1$ is a
light-cone direction.

A related proposal for Matrix theory on ${\bf T}^9/{\bf Z}_2$ was
put forward in \cite{rey}. The gauge group considered there was
$SO(32)$. (It is not possible to take arbitrary $N$ in this case,
due to charge conservation in $1+1$ dimensions.) The authors were
able to show via parton scattering that  the charge of the
orientifold was $-{1\over 32}$ , in agreement with our results.

\vspace{1cm}
\noindent {\Large \bf{Acknowledgments}}
\vspace{0.5cm}

We would like to thank T. Banks, M. Douglas, V. Kac and J. Polchinski for
discussions.

\noindent A.H. is supported in part by the DOE under grant
no. DE-FC02-94ER40818, by an A. P. Sloan Foundation Fellowship and
by a DOE OJI award.
A. R. is supported by DOE grant DE-FG02-96ER40959.
This research was supported in part by the
National Science Foundation under grant no. PHY94-07194.
This work was partly done at the ITP at Santa Barbara during the program on 
Supersymmetric Gauge Dynamics and String Theory.

\end{document}